\documentclass{an}
\usepackage{graphicx}
\usepackage{amssymb}
\usepackage{times}
\usepackage{fancyhdr}
\begin{document}

\title{Spitzer Discovery of Very Low Luminosity Objects}

\author{J.\ Kauffmann\inst{1} \and F.\ Bertoldi\inst{2} \and N.J.\
  Evans II\inst{3} \and the c2d Collaboration}
\institute{
  Max-Planck-Institut f\"ur Radioastronomie, Auf dem H\"ugel
  69, D-53121 Bonn, Germany
  \and
  Radioastronomisches Institut der Universit\"at Bonn, Auf dem H\"ugel
  71, D-53121 Bonn, Germany
  \and
  Department of Astronomy, University of Texas at Austin, 1 University
  Station C1400, Austin, TX 78712-0259
}

\date{Received; accepted; published online}

\abstract{The Spitzer Space Telescope allows for the first time to
  search systematically for very low luminosity ($\lesssim 0.1 \,
  L_{\odot}$) objects (VeLLOs) associated with dense molecular cores.
  They may be the first candidate Class~0
  sources with sub-stellar masses.
  We describe such a source in the dense molecular core L1148. VeLLO
  natal cores  show properties that are unusual for star-forming
  cores.
  The low luminosity and in some cases the lack of prominent
  outflow could be the result of the small 
  gas supply near the VeLLO.
   \keywords{stars: formation --- ISM: globules --- ISM: individual (L1148)} }

\correspondence{jkauffma@mpifr-bonn.mpg.de}

\maketitle

\section{Introduction}
The ``revealed'' phase of early stellar evolution was discovered and
classified in the late 1980's (Lada 1987), but it took some years
until objects in the ``obscured'' phase of star-formation were found
and classified (Andr\'e, Ward-Thompson \& Barsony 1993). Such Class~0
protostars are so deeply embedded in their natal dense cores that they
could not be detected with contemporary near- and mid-infrared
detectors. They are believed to be in the main accretion phase of star
formation (Andr\'e, Ward-Thompson \& Barsony 2000),
drive collimated bipolar outflows (Andr\'e et al.\ 2000) and
can show maser emission (Furuya et al.\ 2001).

While the first Class~0 sources were studied, also the first brown
dwarfs (BDs) and extrasolar planets were discovered (Nakajima et
al.\ 1995). Many BDs are not orbiting a stellar mass primary
(Oppenheimer, Kulkarni \&
Stauffer 2000). There is no proof that these BDs were ejected
from forming protostellar systems. The large BD binary fraction
of $\gtrsim 15\%$ (Mart\'in et al.\ 2003, Pinfield et al.\ 2003) is
furthermore hardly explained by ejection scenarios (Whitworth \&
Goodwin 2005). It thus appears plausible
that at least some of the single brown dwarfs may have formed by
processes similar to those by which normal stars form, i.e. the
collapse of a dense core onto a compact object. Sub-stellar mass
compact objects in the process of formation would appear as
low-luminosity ($L_{\mathrm{bol}} \ll 1 \, L_{\odot}$) Class~0
sources.

It is a challenge to search and study accreting very low luminosity
objects (VeLLOs) that may turn into sub-stellar mass brown dwarfs.
While a first such source (IRAM04191) was found serendipitously
(Andr\'e, Motte \& Bacmann 1999), a systematic search for such objects
has become possible only with the Spitzer Space Telescope (SST).  Our
SST Legacy Program ``From Molecular Cores to Planet Forming Disks''
(c2d; Evans et al.\ 2003), which images about 90 isolated dense cores
and 5 large molecular clouds (Huard et al., this volume), is tailored
to find such VeLLOs. The source L1014-IRS (Young et al.\ 2004) was the
first low luminosity object discovered in our survey. We here present
another candidate for an accreting sub-stellar object, L1148-IRS.

\section{The VeLLO Candidate L1148-IRS\label{sec:VeLLOs}}
\subsection{The compact IR Source}
\begin{figure}
\includegraphics[height=\linewidth,angle=-90]{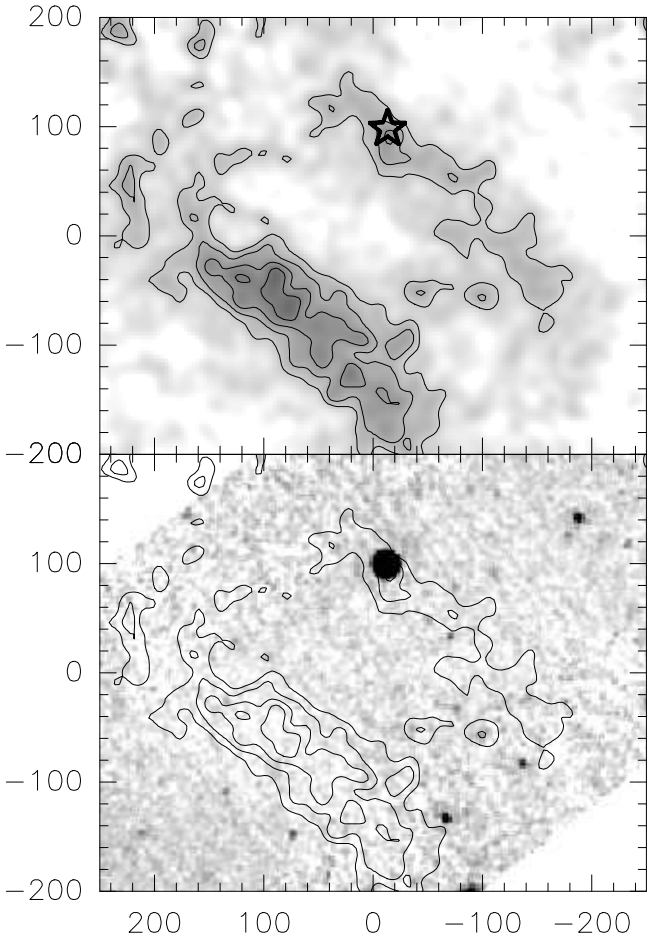}
\caption{\label{fig:L1148_MAMBO_Spitzer} MAMBO dust emission (top) and
  $24 \, \mathrm{\mu m}$ Spitzer map of L1148 (bottom). The bright
  Spitzer source L1148-IRS is marked by a star in the MAMBO
  map. Offsets are in arcseconds.}
\end{figure}

The dense core L1148 is part of the Cepheus flare at a distance of
about $325 \, \mathrm{pc}$ (Straizys et al.\ 1992). We have imaged the
thermal dust emission of L1148 at 1.2~mm wavelength using the
Max-Planck Millimeter Bolometer (MAMBO-2) array camera at the IRAM
30m-telescope on Pico Veleta (Spain). Such wide band continuum
observations map the projected mass distribution, since the dust
emission is nearly proportional to the $\mathrm{H_2}$ column density
and the dust temperature (e.g., Motte, Andr\'e \& Neri 1998). In our
mm-continuum map we found two parallel, faint filaments in the
south-east and north-west of L1148 (Fig.\ 
\ref{fig:L1148_MAMBO_Spitzer}). Assuming dust temperatures of
$10 ~ \rm K$ we estimate their masses to 11 and 5 $M_{\odot}$ and
their peak visual extinctions to 7 and 5 mag, respectively.

The dust emission peak in the north-western filament is associated
with a bright Spitzer source at $24 ~ \mu \mathrm{m}$ and
$70 ~ \mu \mathrm{m}$ (Fig.\ \ref{fig:L1148_MAMBO_Spitzer}), named
L1148-IRS. From its appearance in the IRAS Faint Source Catalog Kun
(1998) noted it as a potential low-luminosity protostar. A spatial,
physical association of the dust emission peak and L1148-IRS is
suggested by the $6 \arcsec$ proximity. The probability of a
chance alignment within $20 \arcsec$ of the dust peak (about
twice the separation of dust peak and L1148-IRS) and a background
galaxy of comparable infrared brightness is very low (1:7700 at
$24 ~ \mu \mathrm{m}$, 1:3400 at $70 ~ \mu \mathrm{m}$).
Because L1148 is way above the galactic plane
($z \approx 80 ~ \mathrm{pc}$ at $b = +15^{\circ}$) a chance
alignment with a background protostar is also unlikely. Therefore
L1148-IRS is most probably an infrared source associated with the L1148
dense core.

Using the IRAM 30m-telescope we have unsuccessfully searched for a CO
outflow from L1148-IRS, which would have clearly identified L1148-IRS
as an accreting protostar. This does not exclude the presence of a
small-scale flow, which in the VeLLO L1014-IRS could only be detected
with an interferometer (Sec.\ \ref{sec:IRAM_L1014}).

\begin{figure}
\includegraphics[width=\linewidth,angle=0]{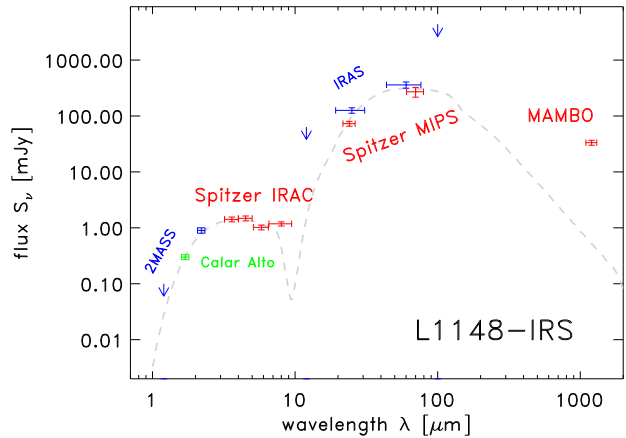}
\caption{\label{fig:L1148_SED} The spectral energy distribution of
  L1148-IRS. The dashed line shows our model SED for the compact
  emission from an embedded protostar (which does
  not apply to the MAMBO data point).}
\end{figure}

With the available archival 2MASS and IRAS data and our own
OMEGA-Cass.\ (Calar Alto 3.5m-telescope) near-IR observations, in
Fig.\ \ref{fig:L1148_SED} we plot the observed spectral energy
distribution (SED) of L1148-IRS. For
$\lambda \lesssim 100 \, \mathrm{\mu m}$ the source L1148-IRS is
compact (smaller than a few arcsec) and the fluxes listed hold for
this compact source. For $\lambda \gtrsim 100 \, \mathrm{\mu m}$ the
source is extended (several arcmin at
$\lambda = 1 \, 200 \, \mathrm{\mu m}$) and the fluxes hold for an
aperture of $12 \farcs 9 = 4 \, 200 ~ \rm AU$ radius, to be consistent
with previous work (Andr\'e et al.\ (1999).

To estimate the bolometric luminosity of the embedded star we
integrated the flux density over the observed SED using a modified
version (at $\lambda \gtrsim 200 ~ \rm \mu m$ we assume the SED to
follow a greybody of temperature $10 ~ {\rm to} ~ 15 ~ \rm K$ and dust
opacity $\kappa \propto \lambda^{-2}$ instead of a powerlaw) of the
method of Chen et al.\ (1995). The integration yields
$L_{\mathrm{bol}} = 0.1 ~ \mathrm{to} ~ 0.2 \, L_{\odot}$
(out of which about 30 to 70\% are
due to heating by the interstellar radiation field [ISRF]), a
submillimeter-to-bolometric luminosity ratio
$L_{>350 \mu \mathrm{m}} / L_{\mathrm{bol}} \ge 0.05$, and a
bolometric temperature $T_{\mathrm{bol}} \le 140 ~ \mathrm{K}$.
As $L_{>350 \mu \mathrm{m}} / L_{\mathrm{bol}} \gg 0.005$
L1148-IRS qualifies as a Class~0 protostar
(Andr\'e et al.\ 2000).

We tried to model the observed SED with an embedded young
stellar object (Fig.\ \ref{fig:L1148_SED}). The continuum 
radiative transfer was calculated\footnote{The code was kindly provided
  by E.\ Kr\"ugel.} for a star of luminosity $L$ and
temperature $T$,
surrounded by a spherical core with a power-law density profile:
$\varrho \propto r^{- \alpha}$ with $\alpha = 3/2 ~ {\rm to} ~ 2$
(Shu 1977) between radius $r_{\rm in}$ and $r_{\rm trans}$,
$\alpha = 0$ between $r_{\rm trans}$ and $r_{\rm out}$,
$\varrho=0$ for $r < r_{\rm in}$ and $r > r_{\rm out}$.
While $r_{\rm in}$ and $r_{\rm trans}$ are free
parameters, the outer radius, $r_{\rm out}$, and density are
constrained by the extended dust emission seen in the MAMBO maps. The
core is exposed to the ISRF. We take the differential observing
technique (e.g., a ``background intensity'' is subtracted from a
Spitzer image before a flux measurement) into account when modelling
the individual flux densities.  This produces different model SEDs for
the compact ($<2$ arcsec at $\lambda \lesssim 100 \, \rm \mu m$) and the
extended emission ($>2$ arcsec at
$\lambda \gtrsim 100 \, \rm \mu m$). The main result from fitting our
models is that independent of our choice of values for the free
parameters, the stellar luminosity and temperature can be
constrained to $L = 0.05 ~ \mathrm{to} ~ 0.15  L_{\odot}$
and $T = 2 \, 000 ~ \mathrm{to} ~ 4 \, 000 \mathrm{K}$.

In a detailed comparison we notice a discrepancy between the predicted
and observed flux densities in the $8 ~ \mu \textrm{m}$ Spitzer band
(Fig.\ \ref{fig:L1148_SED}), which may be dominated by silicate
absorption and emission. Furthermore, our model fit implies an inner
radius, $r_{\rm in}$, that is much larger than the dust sublimation
radius (the dust in our model has temperatures $\lesssim 100 ~ \rm K$,
where dust sublimates at $\approx 1 \, 500 ~ \rm K$), possibly
suggesting a more complicated geometry including the presence of
a disk and cavities. The presence of significant
amounts of hot ($\gg 100 ~ \rm K$) dust, in which case we would
observe a significant slope in the IRAC bands, can be excluded.

Following Young et al.\ (2004) we derive an upper limit to the
protostar mass, $M_{\ast}$, using the constraint that the accretion
luminosity, $L_{\mathrm{accr}} = G M_{\ast} \dot{M} / R_{\ast}$ 
is smaller than the total luminosity, $L_{\mathrm{accr}} < L$. 
Here $G$ is the constant of gravity, $R_{\ast}$ is the protostar
radius, and $\dot{M}$ the accretion rate. Using estimates for
$R_{\ast}$ and $\dot{M}$ from Young et
al.\ (2004), we find $M_{\ast} \ll 0.1 \, M_{\odot}$.

\subsection{The Natal Dense Core}
\begin{figure}
\includegraphics[height=\linewidth,angle=-90]{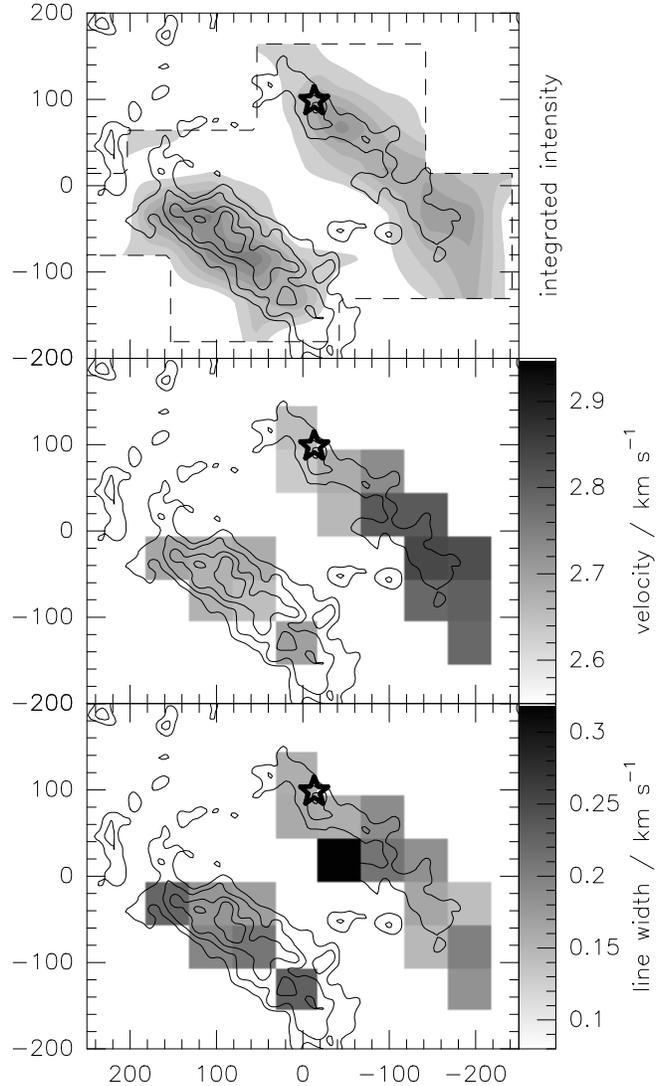}
\caption{\label{fig:L1148_CCS} CCS $2_1$-$1_0$ integrated intensity,
  centroid velocity, and velocity dispersion.
  Overlayed contours are dust continuum intensity. The
  position of L1148-IRS is marked by a star. Offsets are arcseconds.}
\end{figure}

Our Effelsberg 100m telescope maps of L1148 
in the $2_1$-$1_0$ transition of CCS probe the dense core on large
scales (Fig.\ \ref{fig:L1148_CCS}). 
The velocity-integrated CCS intensity map
reveals two filaments associated with the dust emission filaments. 
The CCS centroid velocity map shows a jump in
velocity about $70 \arcsec$ south-west of L1148-IRS. Since the
velocity field is smooth elsewhere,  the velocity-discontinuity
suggests two neighbouring, possibly overlapping dense cores, separated in
velocity by $0.1 ~ \mathrm{km ~ s^{-1}}$. The linewidth is
$<0.2 ~ \mathrm{km ~ s^{-1}}$ in the north-western filament, except where
the gradient of the centroid velocity is large, possibly
due to overlapping velocity components.

Our molecular line observations with the IRAM 30m-telescope probe the
dense core on smaller scales. The $\mathrm{C^{18}O}$ 2-1 emission,
combined with the dust emission implies a CO depletion of
$f_{\mathrm{d}} = 8$ (Crapsi et al.\ 2005a); the abundance of CO in
the gas phase appears to be reduced by a factor 8 relative to the
``standard'' abundance, probably through freeze-out onto dust grains
during the evolution of the core. Some of the observed
$\mathrm{C^{18}O}$ emission could come from velocity components
unrelated to the dust peak, in which case $f_{\mathrm{d}}$ for the
dense core would be underestimated.

\begin{figure}
\includegraphics[height=0.8\linewidth,angle=-90]{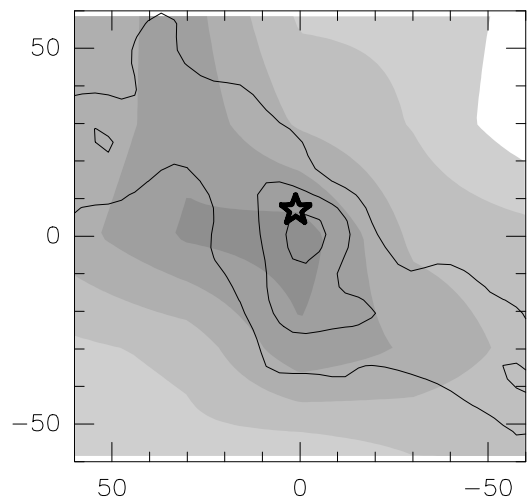}
\caption{\label{fig:L1148_N2H+} Dense gas near L1148-IRS as probed in
  the $\mathrm{N_2H^+}$ (1-0) transition. The position of L1148-IRS is
  marked by a star. Offsets are in seconds of arc.}
\end{figure}

Our map of the $\mathrm{N_2H^+}$ 1-0 emission shows that the dense gas
is closely associated with L1148-IRS (Fig.\ \ref{fig:L1148_N2H+}),
further supporting the picture that the Spitzer source is physically
associated with the L1148 dense core.



%

\section{The confirmed Protostars IRAM04191 and L1014-IRS \label{sec:IRAM_L1014}}
Besides L1148-IRS, two other low-luminosity protostars have been studied in
detail. 

IRAM04191 was the first known protostar with a very low
luminosity ($L \approx 0.1 \, L_{\odot}$) and therefore mass (Andr\'e
et al.\ 1999). It was found through its prominent CO outflow, that is
associated with a VLA cm radio source and a dust emission
peak. A near- and far-infrared counterpart was recently identified in
our Spitzer data (Dunham et al.\ 2005).

The VeLLO L1014-IRS was discovered through the association
of an unusually bright Spitzer source with a dust emission peak (Young
et al. 2004). It has a low luminosity of 
$\sim 0.1 \, L_{\odot}$. The physical association of the Spitzer source with
the L1014 dense core has been established by the detection of an infrared
outflow cone (Huard et al.\ 2005; see also Huard et
al., this volume) and the discovery of a small-scale CO outflow with 
the SMA interferometer (Bourke et al.\ 2005).

\section{The Physics of Very Low Luminosity Objects\label{sec:formation}}

Compared with previously discovered Class~0 protostars, VeLLOs appear to be
surrounded by small amounts of molecular gas.  Motte \& Andr\'e (2001) suggest
a comparison based on the mass within a projected radius of $4200 ~
\mathrm{AU}$ from the protostar, $M^{4200 ~ \mathrm{AU}}_{\mathrm{env}}$.

In the Class~0 source sample of Motte \& Andr\'e (2001),
IRAM04191 has the smallest inner envelope mass (and luminosity), and the
masses for L1014 and L1148 are even smaller than that (Tab.\
\ref{tab:VeLLOs}). It is suggestive that for VeLLOs the low luminosity and
lack of outflow activity somehow correlate with the low envelope
mass. However, such conclusion is yet based on too small
a sample.

\begin{table}
  \caption{\label{tab:VeLLOs} Properties of VeLLOs (top) and their
    natal dense cores (bottom) as observed with single dish telescopes. 
    Dense cores are compared with criteria for ``evolved'' cores
    from Crapsi et al.\ (2005b); $N$ is column density. A plus
    indicates that a criterion is met, a minus that it is not, a
    question mark  that the data is insufficient to decide. Data are
    from Motte \& Andr\'e (2001),
    Belloche et al.\ (2002), Crapsi et al.\ (2005b), and this work.}
  \begin{tabular}{lccc}
    \hline \rule{0ex}{3Ex}
    Property & IRAM04191 & L1014 & L1148 \vspace{1ex}\\ \hline
    \multicolumn{4}{l}{\itshape protostar properties:}\\
    prominent outflow & + & $-$ & $-$\\
    $M^{4200 ~ \mathrm{AU}}_{\mathrm{env}} / M_{\odot}$ &
    $0.45$ & $0.27$ & $0.11$ \vspace{1ex}\\
    \multicolumn{4}{l}{\itshape dense core properties:}\\
    $f_{\mathrm{d}} > 10$ & ? & $-$ & $-$\\
    $N({\mathrm{N_2D^+}}) / N({\mathrm{N_2H^+}})$ & ? & + & ?\\
    $ \ge 0.1$\\
    $N({\mathrm{N_2H^+}})$ & + & $-$ & $-$\\
    $\ge 8.5 \cdot 10^{12} ~ \mathrm{cm^{-2}}$\\
    contr.\ motions, infall & + & $-$ & +\\ \hline
  \end{tabular}
\end{table}

VeLLOs are found in cores that
were thought not to be able to form stars.
Crapsi et al.\ (2005a) suggested criteria for properties of 
starless cores just before the onset of star-formation.
According to this, cores must be exceptionally dense 
and chemically evolved, e.g., they show high degrees of
deuteration and depletion. Table 
\ref{tab:VeLLOs} shows that the star-forming cores L1014 and L1148 meet only
some of Crapsi's criteria. Since these were put forward on the basis of single-dish
observations, they may be insufficient to probe the VeLLO natal dense
cores on those scales relevant for the star-formation process.

We remark (as Andr\'e et al.\ 1999 did for
IRAM04191 and Young et al. 2004 did for L1014) that VeLLOs are
embedded in dense cores with sufficiently large masses that they could
in principle grow to stellar masses ($>0.1 M_\odot$), and thus do not
necessarily remain sub-stellar objects.


\section{Summary\label{sec:summary}}
The Spitzer Space Telescope allows for the first time a systematic
search for very low luminosity ($L \approx 0.1 \, L_{\odot}$)
embedded, compact objects (VeLLOs), 
accreting Class~0 sources of apparently sub-stellar
mass. With L1148-IRS we here present a further candidate for such an object. 


The presently known VeLLOs have a very low inner envelope mass. This could 
relate to their low luminosities and -- in the case of L1014-IRS and
L1148-IRS -- to their low outflow activity. Two of the three 
sufficiently studied VeLLO natal cores
(L1014-IRS and L1148-IRS) have properties that do not meet the
proposed criteria for starless cores before the onset of
star-formation. They appear neither very dense nor
chemically evolved when probed with single-dish telescopes. 
VeLLOs seem to form in clouds that were not expected to form stars.



\acknowledgements We thank E.\ Kr\"ugel, T.\ Bourke, P.\ Myers, C.W.\ 
Lee, T.\ Huard, and P.\ Andr\'e for insightful discussions, and the
Instituto de Astrof\'isica de Canarias (IAC) for providing a
very stimulating atmosphere at the ULMSF05 workshop.


\end{document}